# A new kind of boundary layer flow due to Lorentz forces

Asterios Pantokratoras
Associate Professor of Fluid Mechanics
School of Engineering, Democritus University of Thrace,
67100 Xanthi – Greece

e-mail:apantokr@civil.duth.gr

Abstract
In this paper we present a new kind of boundary layer flow produced by an electromagnetic Lorentz force which acts parallel to the plate in an electrically conductive fluid. The plate is motionless and the ambient fluid stagnant. This flow is equivalent to the classical free convective flow along a vertical plate. The boundary layer equations are transformed to non-dimensional form and a new dimensionless number is introduced which is equivalent to the Grashof number. The transformed boundary layer equations are solved with the finite difference method and the presented results include values of the friction coefficient and velocity profiles.

INTRODUCTION

Magnetohydrodynamis is the study of the interaction between magnetic fields and moving, conducting fluids. (Davidson, 2006). In this case a force is produced inside the fluid which is proportional to fluid velocity and this force always opposes the flow. Another way to produce a force inside a flowing fluid, not known widely, is the application of an externally applied magnetic as well as an externally applied electric field. (see figure 1). This force is called Lorentz force and can be generated by a stripwise arrangement of flush mounted electrodes and permanent magnets of alternating polarity and magnetization. The Lorentz force which acts parallel to the plate can either assist or oppose the flow. The idea of using a Lorentz force to stabilize a boundary layer flow over a flat plate belongs probably to Gailitis and Lielausis (1961). The idea of using the Lorentz force for flow control by Gailitis and Lielausis was later abandoned and only recently attracted new attention (Henoch and Stace, 1995, Crawford and Karniadakis, 1997, Berger et al. 2000, Spong et al. 2005). In addition, in last years much investigation on flow control using the Lorentz force is being conducted at the Rossendorf Institute and at the Institute for Aerospace Engineering in Dresden, Germany (Posdziech and Grundmann, 2001, Weier and Gerbeth, 2004, Weier, 2005, Mutschke et al. 2006, Albrecht and Grundmann, 2006).

Although much investigation has been done in different fields simple problems, including a Lorentz force, have not been investigated until now. The purpose of the present paper is to analyze the boundary layer flow produced by a Lorentz force acting parallel to the plate while the plate and the ambient fluid are at rest. This problem is "equivalent" to the classical free convection along a vertical plate due to buoyancy forces. In our case, the role of buoyancy forces play the Lorentz forces while there is no need for the plate to be vertical, because the Lorentz forces are independent of gravity. It is a problem of "free convection" independent of gravity.

THE MATHEMATICAL MODEL

Consider the flow along a plate with $u$ and $v$ denoting respectively the velocity components in the $x$ and $y$ direction, where $x$ is the coordinate along the plate and $y$ is the coordinate perpendicular to $x$. It is assumed that in the flow over the plate a Lorentz force exists. For steady, two-dimensional flow the boundary layer equations with constant fluid properties are (Tsinober and Shtern, 1967, Weier, 2005)

continuity equation: $\quad\dfrac{\partial u}{\partial x}+\dfrac{\partial v}{\partial y}=0 \quad$ (1)

momentum equation: $\quad u\dfrac{\partial u}{\partial x}+v\dfrac{\partial u}{\partial y}=\nu\dfrac{\partial^2 u}{\partial y^2}+\dfrac{\pi j_0 M_0}{8\rho}\exp(-\dfrac{\pi}{a}y) \quad$ (2)

where $\nu$ is the fluid kinematic viscosity, $j_0$ (A/m$^2$) is the applied current density in the electrodes, $M_0$ (Tesla) is the magnetization of the permanent magnets, $a$ is the width of magnets and electrodes and $\rho$ is the fluid density. The last term in the momentum equation is due to the Lorentz force, decreases exponentially with $y$ and is independent of the flow.
The boundary conditions are:

at $y = 0$: $\quad u = 0, \; v = 0$ $\quad$ (3)
as $y \to \infty$ $\quad u = 0$ $\quad$ (4)

We introduce the following dimensionless quantities

$$Pa = \dfrac{\pi j_0 M_0 a^3}{8\rho\nu^2} \quad (5)$$

$$x^* = \dfrac{x}{a} \quad (6)$$

$$y^* = \dfrac{y}{a} Pa^{1/4} \quad (7)$$

$$u^* = \dfrac{ua}{\nu} Pa^{-1/2} \quad (8)$$

$$v^* = \dfrac{va}{\nu} Pa^{-1/4} \quad (9)$$

Using the above dimensionless quantities equations (1)-(2) take the following non-dimensional form continuity equation: $\quad\dfrac{\partial u^*}{\partial x^*}+\dfrac{\partial v^*}{\partial y^*}=0 \quad$ (10)

momentum equation: $$u^* \frac{\partial u^*}{\partial x^*} + v^* \frac{\partial u^*}{\partial y^*} = \frac{\partial^2 u^*}{\partial y^{*2}} + \exp(-\pi y^* Pa^{-1/4}) \quad (11)$$

From the above equations it is clear that the only governing parameter of this problem is the quantity *Pa*.

In the classical problem of free convection along a vertical motionless isothermal plate, with temperature $T_w$, situated in a calm fluid with temperature $T_\infty$, the governing parameter is the well known Grashof number defined as (Jaluria, 1980, Bejan, 1995, White, 2006)

$$Gr = \frac{g\beta(T_w - T_\infty)l^3}{\nu^2} \quad (12)$$

where *β* is the volumetric expansion coefficient of the fluid and *l* is the characteristic length of the flow. The Grashof number expresses the balance between the buoyant forces to viscous forces and was introduced in fluid mechanics by the German engineer Franz Grashof (1826-1893). Although the Grashof number is widely used in heat transfer and fluid mechanics, the man for whom the grouping was named is not familiar to workers in the field. Historical and bibliographical information on this matter is given by Sanders and Holman (1972).

Comparing equations (5) and (12) we see that the new dimensionless number *Pa* is equivalent to the Grashof number with characteristic flow length the width of magnets or the width of electrodes (see figure 1). The quantity *Pa*, which expresses the balance between the electromagnetic body forces to viscous forces, is introduced here for the first time and may be called the Pantokratoras number.

After the introduction of *Pa* and the transformation we solved the above equations (10)-(11) using the finite difference method of Patankar (1980). The solution procedure starts at the plate leading edge ($x^*=0$) and marches along the plate. At each downstream position the discretized equation (11) is solved using the tridiagonal matrix algorithm (TDMA). The cross-stream velocities *v* were obtained from the continuity equation. The forward step size $\Delta x^*$ was 0.001 and the lateral grid cells were 500. The results are grid independent. The parabolic (marching) solution procedure is a well known solution method, has been used extensively in the literature and has been included in widely used textbooks (Jaluria and Torrance, 1986, page 173 and White 2006, page 276). A detailed description of the solution procedure may be found in Pantokratoras (2002).

RESULTS AND DISCUSSION

The results of the present work are shown in table 1 and in figures 2, 3 and 4. Table 1 contains values of the dimensionless skin friction defined as

$$C_f = \left.\frac{\partial u^*}{\partial y^*}\right|_{y^*=0} \tag{13}$$

Table 1. Friction coefficients for different values of $x^*$ and $Pa$

| $x^*$ | Pa=1 | Pa=10 | Pa=100 | Pa=1000 |
|---|---|---|---|---|
| 1 | 0.2922 | 0.4905 | 0.7225 | 0.9636 |
| 5 | 0.3092 | 0.5299 | 0.8414 | 1.2016 |
| 10 | 0.3126 | 0.5406 | 0.8821 | 1.3015 |
| 50 | 0.3147 | 0.5549 | 0.9456 | 1.5010 |
| 100 | 0.3172 | 0.5586 | 0.9630 | 1.5705 |
| 200 | 0.3181 | 0.5617 | 0.9773 | 1.6288 |
| 300 | 0.3187 | 0.5635 | 0.9843 | 1.6584 |
| 500 | 0.3197 | 0.5660 | 0.9927 | 1.6917 |
| 1000 | 0.3219 | 0.5706 | 1.0051 | 1.7337 |

The results of table 1 are shown in figure 2. We see that the friction coefficient increases with increasing $Pa$ and $x^*$. This is in accordance with the velocity profiles, shown in figures 3 and 4, where we see that velocity increases with increasing $Pa$ and $x^*$. However, in figure 4, although the velocity increases with increasing $Pa$, the velocity width decreases at the same $x^*$.

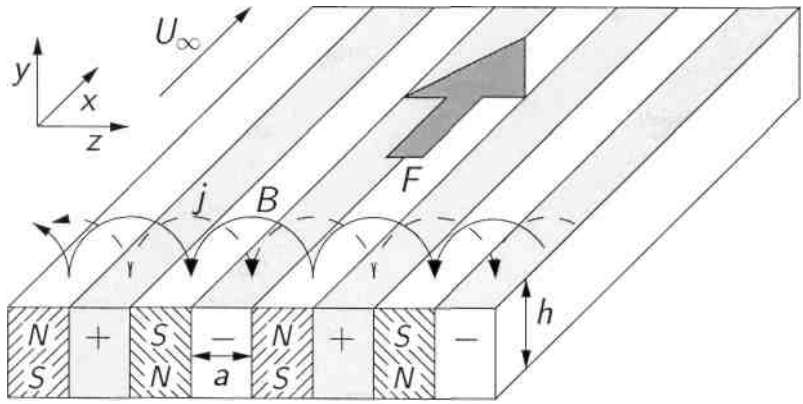

Figure 1. Arrangement of electrodes and magnets for the creation of a Lorentz force F in the flow along a flat plate (Weier, 2005).

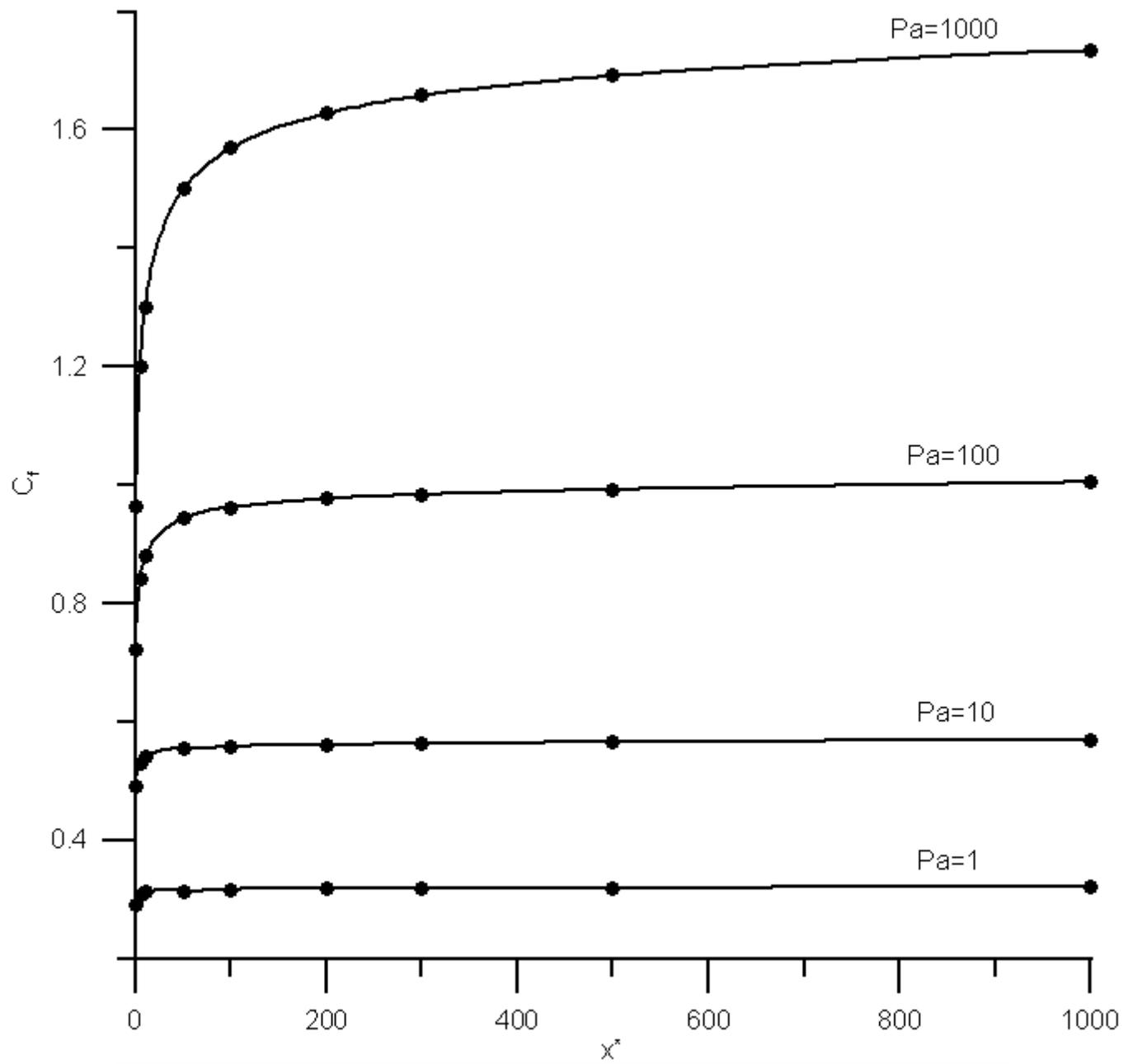

Figure 2. Variation of dimensionless friction coefficient with dimensionless distance along the plate.

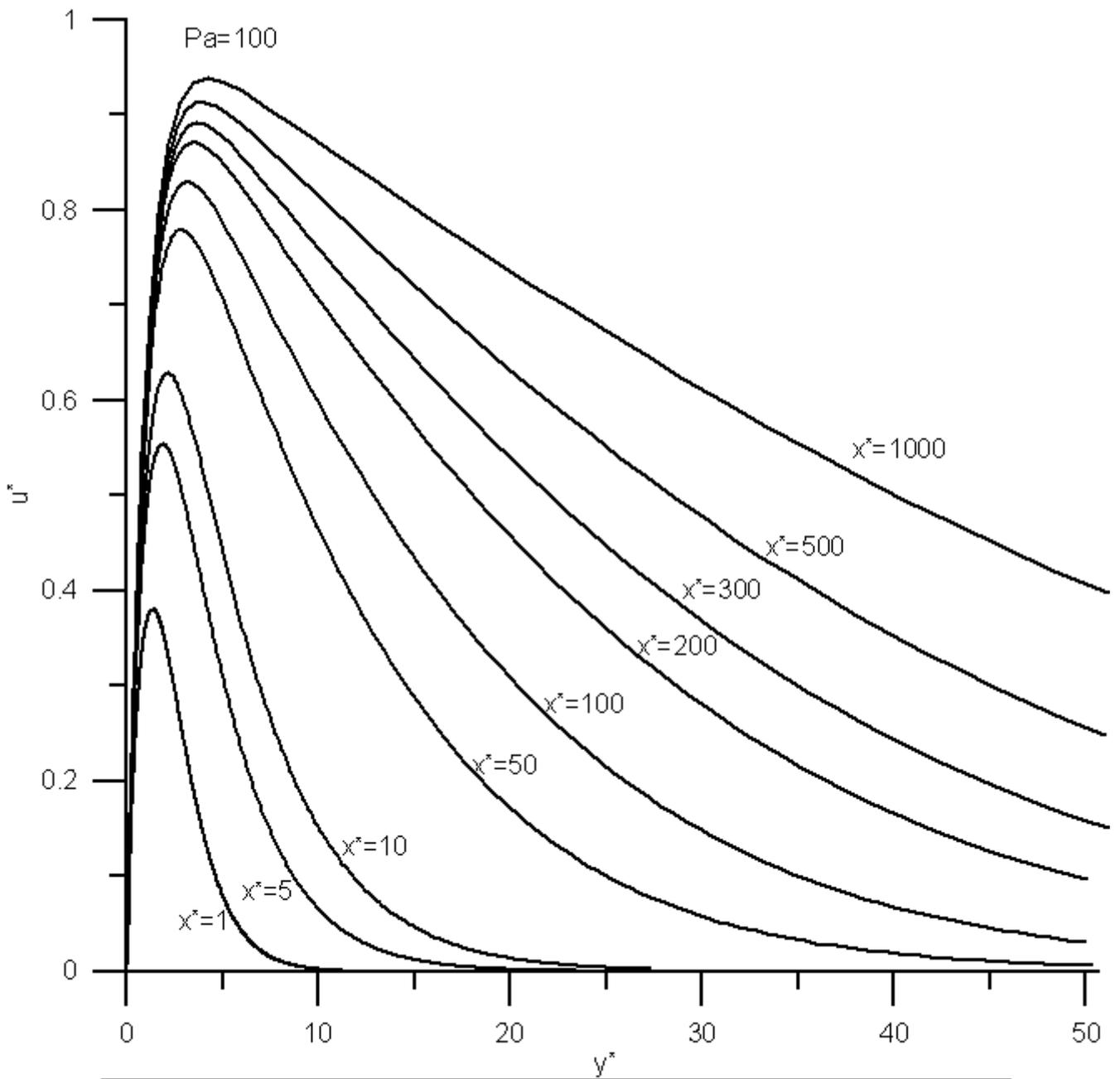

Figure 3. Velocity profiles for Pa=100 and different dimensionless distances from the plate edge.

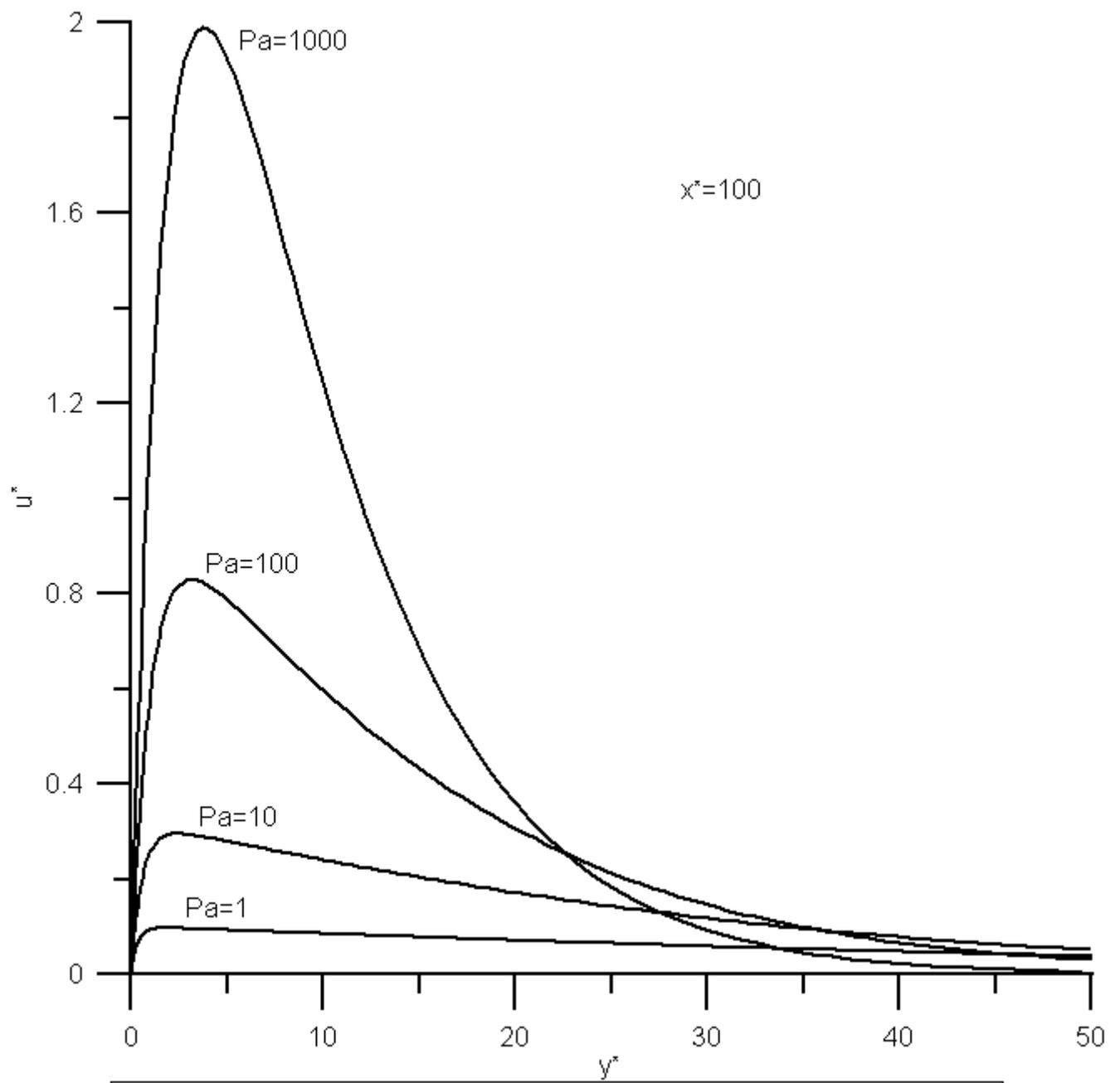

Figure 4. Velocity profiles for x*=100 and different values of Pa parameter.